\documentclass[12pt]{article}      
\usepackage{epsfig}     
\newcommand{\eq}{\begin{equation}}      
\newcommand{\eqx}{\end{equation}}      
\newcommand{\eqn}{\begin{eqnarray}}      
\newcommand{\eqnx}{\end{eqnarray}}      
\newcommand{\dt}{\Delta}

\newcommand{\gp}{g_1^{\gamma}}      
\title{{\bf QCD predictions for spin dependent photonic structure function      
$g_1^{\gamma}(x,Q^2)$ in the low $x$ region       
of future linear colliders}}      
\author{{\sc J.~Kwieci\'nski} \footnote{e-mail:jkwiecin@solaris.ifj.edu.pl},      
        {\sc B. Ziaja}$^{*\,,\,}$ \footnote{e-mail:beataz@solaris.ifj.edu.pl}\\ \\      
         $^*$ \it Department of Theoretical Physics,\\      
                 \it H.~Niewodnicza\'nski Institute of Nuclear Physics,\\      
                 \it 31-342 Cracow, Poland\\ \\      
         $^\dag$ \it High Energy Physics,\\      
                 \it Uppsala University,\\      
                 \it P.O. Box 535,\\      
                 \it S-75121 Uppsala, Sweden }      
\date{}      
\begin{document}      
\maketitle      
\begin{abstract}      
Spin dependent structure function $g_1^{\gamma}(x,Q^2)$ of the polarised photon is analysed within the formalism based upon the       
unintegrated spin dependent parton distributions incorporating the LO Altarelli-Parisi evolution and       
the double $ln^2(1/x)$ resummation at low values of Bjorken parameter $x$.        
We analyse the effects of the double $ln^2(1/x)$ resummation on the       
behaviour of $g_1^{\gamma}(x,Q^2)$ in the low $x$ region which may be        
accessible in future linear $e^+e^-$ and $e\gamma$ colliders.  Sensitivity of       
the predictions on the possible nonperturbative gluon content of the polarised photons is       
analysed.  Predictions for spin dependent gluon distribution $\Delta g^{\gamma}      
(x,Q^2)$ are also given.       
\end{abstract}      
\section{Introduction}      
The small $x$ behaviour of the spin dependent structure functions, where       
$x$ is the Bjorken parameter,    
may be influenced in QCD by the novel effects       
coming from  the double $ln^2(1/x)$ resummation \cite{BARTNS,BARTS,RG,      
JAP,BBJK,BZIAJA}. These effects  generate more singular behaviour in the limit    
$x \rightarrow 0$ than that    
given by LO (or NLO) Altarelli-Parisi evolution equations with    
non-singular input.    
The double $ln^2(1/x)$ resummation effects have been quantitatively    
analysed for  spin dependent       
structure function $g_1(x,Q^2)$ of the nucleon \cite{BZIAJA},  and it has been found that       
they can in principle significantly affect the structure function in the small $x$ region which       
can be probed at  possible polarised HERA       
measurements  \cite{ALBERT,DESYW}.  The formalism developed in \cite{BZIAJA}  
contains important subleading $ln^2(1/x)$ effects  which follow from  
including the complete splitting functions  and the running QCD coupling.    
It may however be still possible that other subleading  terms can appreciably 
reduce the magnitude of the leading double $ln^2(1/x)$ resummation \cite{JAP}.\\    
        
In this paper we would like to investigate        
spin structure function $g_1^{\gamma}$ of the polarised photon in the low $x$       
region which could become accessible in future linear $e^+ e^-$ or $e\gamma$       
linear colliders \cite{STRATMANN1,STRATMANNL,STRATMANN2,STRATMANN3,AVOGT}.      
The $e\gamma$  mode,  i.e. deep inelastic electron (positron) scattering on a     
photon beams obtained through the Compton back-scattering of the  laser photon     
beams \cite{GAMMAC},  would be particularily suitable for probing  the     
photon structure in the region of low values of Bjorken parameter  $x$    
\cite{AVOGT}.       
Possible potential of probing the spin dependent structure functions     
of the photon at both $e^+ e^-$ and $e \gamma$ colliders is discussed in ref.     
\cite{STRATMANN2}.      
The spin dependent parton distributions of polarised photons       
could also be studied through the dijet photoproduction in polarised $ep$ collisions       
at HERA \cite{STRATMANN3}.\\        
      
The content of our paper is as follows: in the next section we discuss       
the double $ln^2(1/x)$ resummation for the case of the spin dependent       
parton distributions and for spin dependent  structure function $g_1^{\gamma}(x,Q^2)$       
of the photon.           
The novel feature of the corresponding integral equations       
for the unintegrated parton distribution functions is the presence of the       
additional contribution to the inhomogeneous terms which is  generated  by       
the point-like coupling of the photon to quarks and antiquarks.      
We  present  analytic  solution of these integral equations  using      
the  approximation of the double $ln^2(1/x)$  resummation by that contribution   
which corresponds to the ladder diagrams.       
In Sec.\ 3 we extend the formalism by including complete leading order (LO)      
Altarelli-Parisi evolution at large values of $x$. We also include the complete      
double $ln^2(1/x)$ resummation by adding the corresponding higher order terms to the kernels      
of the integral equations.  This extension leads to the system       
of integral equations for the unintegrated spin dependent parton distributions       
in the photon  which       
embodies the complete LO Altarelli-Parisi       
evolution at large values of $x$ and the double $ln^2(1/x)$ resummation       
at small $x$.           
In Sec.\ 4 we discuss the solutions of these equations and show predictions       
for  spin dependent structure function $g_1^{\gamma}(x,Q^2)$ and       
for the spin dependent gluon distributions in the photon.  We study       
sensitivity of the predictions upon the assumptions concerning the 
non-perturbative spin dependent gluon       
distributions in the photon.  We analyse two possible scenarios:       
(a) the case in which the non-perturbative part of the quark and gluon       
distributions are neglected and the partonic content of the polarised       
photon is entirely driven by the point-like coupling of the photon       
to quarks and antiquarks and (b) the case in which one introduces in a model       
dependent way the       
non-perturbative gluon distributions in the photon.           
Finally in Sec.\ 5 we give summary of our results.

\section{Double $ln^2(1/x)$ resummation effects for the spin dependent parton distributions       
in the polarised photon}            
Low $x$ behaviour of photonic spin structure function is influenced      
by double logarithmic  $ln^2(1/x)$ contributions i.\ e.\  by those terms      
of the perturbative expansion which correspond to the powers of $ln^2(1/x)$      
at each order of the expansion, similarly      
as for DIS spin dependent structure function $g_1$ of the nucleon \cite{BARTNS,BARTS}.      
In the following we will apply the double $ln^2(1/x)$ resummation scheme      
based on unintegrated parton distributions      
\cite{BBJK,BZIAJA,MAN}. Conventional integrated spin dependent parton      
distributions $\Delta p_l(x,Q^2)$ ($p=q,g$) are related to  unintegrated      
parton distributions $f_l(x^{\prime},k^2)$ in the following way ~:      
\eq      
\Delta p_l(x,Q^2)=\Delta p_l^{(0)}(x)+      
\int_{k_0^2}^{W^2}{dk^2\over k^2}\,f_l(x^{\prime}=x(1+{k^2\over Q^2}),k^2),      
\label{dpi}      
\eqx      
where $\Delta p_l^{(0)}(x)$ is the nonperturbative part of the distribution,    
$k^2$ denotes the transverse momentum    
squared of the probed parton,      
$W^2$ is the total energy in the center of mass $W^2=Q^2\,(\frac{1}{x}-1)$,      
and index $l$ specifies the parton flavour.      
The parameter $k_0^2$ is the infrared cut-off which will be set equal      
to 1 GeV$^2$. The nonperturbative part $\Delta p_l^{(0)}(x)$ can be viewed      
upon as originating from the integration over non-perturbative region      
$k^2<k_0^2$, i.\ e.\      
\eq      
\Delta p_l^{(0)}(x)= \int_{0}^{k_0^2}{dk^2\over k^2}f_l(x,k^2).      
\label{gint0}      
\eqx      
Photonic structure function $g_1^{\gamma}(x,Q^2)$ is related in a standard way      
to the (integrated) parton distributions describing the parton content of      
polarized photon~:      
\eq      
\gp(x,Q^2)=\frac{<e^2>}{2}\,     
[\Delta q_{NS}(x,Q^2)+\Delta q_S(x,Q^2)],      
\label{g11}      
\eqx      
where $N_f$ denotes the number of active flavours ($N_f=3$). For convenience      
we have introduced in (\ref{g11}) the non-singlet and singlet combinations      
of the spin dependent quark and antiquark distributions defined as:      
\eq      
\Delta q_{NS}(x,Q^2)= \sum_{l=1}^{N_f} \left({e_l^2\over\langle e^2 \rangle}      
- 1\right)(\Delta q^{\gamma}_{l}(x,Q^2) + \Delta \bar q^{\gamma}_{l}(x,Q^2)),      
\label{dpns}      
\eqx      
\eq      
\Delta q_S(x,Q^2)= \sum_{l=1}^{N_f}(\Delta q^{\gamma}_{l}(x,Q^2) + \Delta \bar q^{\gamma}_{l}(x,Q^2)),      
\label{fs}      
\eqx      
where $\langle e^m\rangle =       
{1\over N_f}\sum_{l=1}^{N_f}(e_l)^m$.      
      
\begin{figure}[t]      
    \centerline{\epsfig{figure=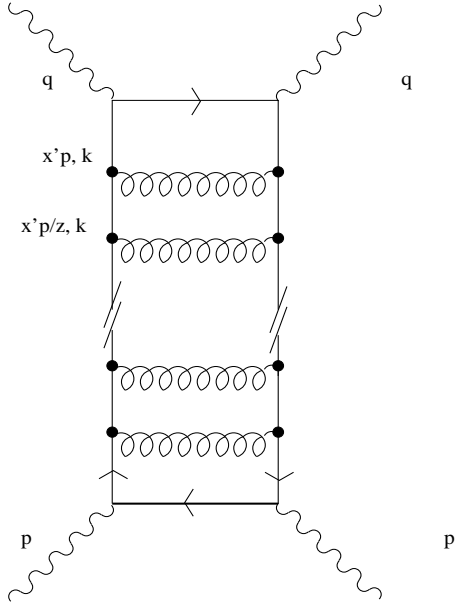,height=8cm,width=5cm}}      
    \caption{{\footnotesize Ladder diagram generating      
     double logarithmic terms in photonic spin structure function      
     $\gp$.}}      
\label{fig.1}      
\end{figure}      

The full contribution to the double $ln^2(1/x)$ resummation comes from      
the ladder diagrams with quark and gluon exchanges along the ladder      
(see e.\ g.\ Fig.\ 1) and the non-ladder bremsstrahlung diagrams \cite{QCD}.      
The latter ones are obtained from the ladder diagrams by adding to them      
soft bremsstrahlung gluons or soft quarks \cite{BARTNS,BARTS,QCD}, and they      
generate the infrared corrections to the ladder contribution.      
      
The relevant region of phase space generating the double      
$ln^2(1/x)$ resummation from  ladder diagrams corresponds      
to ordered $k_n^2/x_n$, where $k_n^2$ and  $x_n$ denote respectively the      
transverse momenta squared and longitudinal momentum fractions of the proton      
carried by  partons  exchanged along the ladder \cite{QED}. It is      
in contrast to the leading order Altarelli - Parisi  evolution alone which corresponds      
to ordered transverse momenta.      
         
The structure of the corresponding integral equations describing unintegrated    
distributions $f_{NS}(x{^\prime},k^2), f_{S}(x{^\prime},k^2)$ and    
$f_g(x{^\prime},k^2)$ in a photon is the same as for the case of the partonic    
structure of a hadron \cite{BZIAJA} which read:       

\eq      
f_{NS}(x^{\prime},k^2)=f^{(0)}_{NS}(x^{\prime},k^2) +      
{\alpha_S \over 2 \pi} \Delta P_{qq}(0)      
\int_{x^{\prime}}^1 {dz\over z}      
\int_{k_0^2}^{k^2/z}      
{dk^{\prime 2}\over k^{\prime 2}}      
f_{NS}\left({x^{\prime}\over z},k^{\prime 2}\right),      
\label{dlxns}      
\eqx      
{\footnotesize      
\eqn      
f_{S}(x^{\prime},k^2)=f^{(0)}_{S}(x^{\prime},k^2)+      
{\alpha_S\over 2 \pi}      
\int_{x^{\prime}}^1 {dz\over z}      
\int_{k_0^2}^{k^2/z}      
{dk^{\prime 2}\over k^{\prime 2}}      
\left[\Delta P_{qq}(0)      
f_{S}\left({x^{\prime}\over z},k^{\prime 2}\right)+      
\Delta P_{qg}(0)      
f_{g}\left({x^{\prime}\over z},k^{\prime 2}\right)\right]\nonumber,      
\eqnx}      
{\footnotesize      
\eqn      
f_{g}(x^{\prime},k^2)=f^{(0)}_{g}(x^{\prime},k^2) +      
{\alpha_S\over 2 \pi}      
\int_{x^{\prime}}^1 {dz\over z}      
\int_{k_0^2}^{k^2/z}      
{dk^{\prime 2}\over k^{\prime 2}}      
\left[\Delta P_{gq}(0)      
f_{S}\left({x^{\prime}\over z},k^{\prime 2}\right)+      
\Delta P_{gg}(0)      
f_{g}\left({x^{\prime}\over z},k^{\prime 2}\right)\right]\nonumber\\      
\label{dlxsg}      
\eqnx}      
with splitting functions $\Delta P_{ij}(0) \equiv \Delta P_{ij}(z=0)$      
equal to~:      
\begin{eqnarray}      
{\bf \dt P(0)} \equiv      
\left( \begin{array}{cc} \Delta P_{qq}(0) & \Delta P_{qg}(0)\\      
\Delta P_{gq}(0) & \Delta P_{gg}(0) \\ \end{array} \right ) =      
\left( \begin{array}{cc}  {{N_C^2-1 \over 2N_C}} & - N_F \\      
                          {{N_C^2-1 \over  N_C}} &  4N_C  \\ \end{array}\right),      
\label{dpij}      
\end{eqnarray}      
\noindent      
where $\alpha_S$ denotes the QCD coupling which at the moment is treated       
as a fixed parameter.       
The variables $k^2$($k^{\prime 2}$) denote the transverse momenta      
squared of the quarks (gluons) exchanged along the ladder.  In the case       
of the parton distributions of a hadron the inhomogeneous       
driving terms  $f^{(0)}_l(x',k^2)$ are entirely determined by non-perturbative       
parts $\Delta p_i^{(0)}(x')$ of the (spin dependent) parton distributions.       
The novel feature of the photon case is the appearance of the additional       
contributions to these inhomogeneous terms         
describing point-like interaction of polarized      
photon with  quarks and antiquarks.   If the non-perturbative parts of the parton       
distributions were neglected then the inhomogeneous terms would be given by:       
\begin{equation}      
f_l^{(0)}(x',k^2) = k^{\gamma}_l(x'),       
\label{ftok}      
\end{equation}      
where  functions $k^{\gamma}_l(x')$ are defined as      
\cite{STRATMANN1,STRATMANNL}:       
\eqn      
k^{\gamma}_{NS}(x)&=&      
\frac{N_C\,N_f}{2\pi}\,\left({\langle e^4\rangle \over \langle e^2\rangle}-   
\langle e^2\rangle \right)\,\,\kappa(x),      
\label{kns}\\      
k^{\gamma }_{S}(x)&=&       
\frac{N_C\,N_f}{2\pi}\,\langle e^2\rangle \,\,\kappa(x),     
\label{ks}\\     
k^{\gamma }_{g}(x)&=&0      
\label{kg}      
\eqnx     
with function $\kappa(x)$ given by     
\begin{equation}      
\kappa(x)=2 \alpha_{em}(x^2-(1-x)^2),      
\label{kappa}      
\end{equation}       
where  $N_C$ denotes number of colours ($N_C=3$), and $\alpha_{em}$ is     
the electromagnetic coupling constant.      
To be precise, in the genuine double $ln^2(1/x)$ approximation       
we should set $f^{(0)}_l(x)=k_l^{\gamma}(0)$.      
      
For the inhomogeneous terms set equal to $k_l^{\gamma}(0)$ the solution(s) of       
equations (\ref{dlxns},\ref{dlxsg}) is found to be:       
\eq      
f_{NS}(x,k^2)= d_{NS}\,I\left(2\sqrt{D_{NS}\,y\,\rho},{y\over \rho}\right),      
\label{lns}      
\eqx      
\eq      
f_{S}(x,k^2)= \frac{x^-d_{S}}{x^--x^+}\,       
I\left(2\sqrt{D_{+}\,y\,\rho},{y\over \rho}\right)-      
              \frac{x^+d_{S}}{x^--x^+}\,      
              I\left(2\sqrt{D_{-}\,y\,\rho},{y\over \rho}\right),      
\label{lq}      
\eqx      
\eq      
f_{g}(x,k^2)= \frac{x^-x^+d_{S}}{x^--x^+}\,I\left(2\sqrt{D_{+}\,y\,\rho},{y\over \rho}      
\right) -      
              \frac{x^+x^-d_{S}}{x^--x^+}\,I\left(2\sqrt{D_{-}\,y\,\rho},{y\over \rho}\right),      
\label{lg}      
\eqx      

\noindent     
where $y=\ln(1/x)$, $\rho= \ln[k^2/(k_0^2x)]$,      
$\bar \alpha_S= \alpha_S/(2\pi)$ and       
\eqn      
d_{NS}&=&-2 \,N_F \,N_C \,\left({\langle e^4 \rangle \over \langle e^2 \rangle} -    
\langle e^2 \rangle \right),\nonumber\\      
d_{S} &=&-2 \,N_f \,N_C \,\langle e^2 \rangle ,\nonumber\\      
D_{NS}&=&\bar \alpha_S \,\lambda_{NS},\nonumber\\      
D_{+} &=&\bar \alpha_S \,\lambda_{+},\nonumber\\      
D_{-} &=&\bar \alpha_S \,\lambda_{-}.     
\eqnx      
The  function $I(2\sqrt{u},v)$       
is defined by       
\begin{equation}      
I(2\sqrt{u},v)=\alpha_{em}[(1-v)I_0(2\sqrt{u}) +       
{v\over \sqrt{u}}I_1(2\sqrt{u})],      
\label{iexp}      
\end{equation}      
where $I_i(z)$ are the modified Bessel functions.        
The coefficients $x^-, x^+$ are defined as~:      
\eq      
x^{\pm}=\frac{\lambda_{\pm}-\Delta P_{qq}}{\Delta P_{qg}},      
\eqx      
while $\lambda_{NS}=\Delta P_{qq}(0)$, and $\lambda_{+}$, $\lambda_{-}$       
correspond to the eigenvalues of matrix ${\bf \dt P(0)}$ (\ref{dpij})      
(cf.\ \cite{BZIAJA}).

In the limit of $x \rightarrow 0$ the dominant term in singlet distribution       
$f_{S}(x,Q^2)$ is given by the first term in eq.\ (\ref{lq})       
which is proportional to function  $I\left(2\sqrt{D_{+}y\rho},{y\over \rho}\right)$, since       
$D_{+} > D_{-}$.  It can be checked, however, that  coefficient $x^-$       
which multiplies this function is smaller by about a factor equal to    
three than coefficient $x^+$  multiplying       
function $I\left(2\sqrt{D_{-}\,y\,\rho},{y\over \rho}\right)$.       
This implies that the       
asymptotic behaviour controlled by the leading term is delayed to the very       
small values of $x$ and that for moderately small values of $x$ the singlet       
quark distributions are dominated by the second term in eq.\ (\ref{lq}).  This effect       
is closely related to the approximation in which possible non-perturbative (i.e.       
hadronic) parts of the (spin dependent) distributions are neglected.  It will       
also be present in the more elaborate and realistic treatment of the distributions       
which will be discussed in the subsequent Sections.  The contribution of the leading eigenvalue       
can be enhanced by  the non-perturbative spin dependent       
(input) gluon distributions in the photon.\\

Since $y=ln(1/x)$ and $\rho= \ln[k^2/(k_0^2x)]$ we find that      
$y\rho \sim ln^2(1/x)$ and $y/\rho \sim 1$ in the small $x$ limit.        
 For large values of $u$ and for fixed $v$ function $I(2\sqrt{u},v)$       
behaves like $exp(2\sqrt{u})$ modulo power corrections.        
This implies the power-law       
behaviour $\sim x^{-2\sqrt{D_{+}}}$ (modulo logarithmic corrections)  of the (singlet) parton       
distributions and of structure function $g_1^{\gamma}(x,Q^2)$.\\      
      
Besides the ladder diagrams contributions the double logarithmic resummation does   
also acquire corrections from non-ladder bremsstrahlung       
contributions.  It has been shown in ref. \cite{BZIAJA} that these   
contributions can be included       
by adding the higher order terms to the kernels of  integral equations      
(\ref{dlxns},\ref{dlxsg}).       
These terms can be obtained from the matrix:       
\begin{equation}      
\Biggl[ \frac{\tilde  {\bf F}_8 }{\omega^2} \Biggr](z)      
\bf G_0\,,       
\label{bmatr}      
\end{equation}     
where $\Biggl[ \frac{\tilde  {\bf F}_8 }{\omega^2} \Biggr](z)$ denote the inverse Mellin transform       
of the octet partial wave matrix (divided by $\omega^2$ ), and the matrix      
$\bf G_0$ reads:     
\eqn     
{\bf G}_0 &=&\left( \begin{array}{cc}  {N_c^2-1 \over 2N_c} & 0  \\     
                                                    0 & N_c  \\ \end{array}      
						    \right).	     
\label{g0}						         
\eqnx     
Following ref. \cite{BZIAJA},       
we shall use Born approximation for the octet matrix which gives:      
\eq     
\Biggl[\frac{\tilde {\bf F}_8^{Born}}{\omega^2}\Biggr](z)=     
4\pi^2 \bar \alpha_S{\bf M}_8 ln^2 (z),     
\label{born}     
\eqx     
where $\bf M_8$ is the splitting functions matrix in colour octet t-channel,     
and it takes the form:     
\eqn     
{\bf M}_8 &=&\left( \begin{array}{cc} -{1 \over 2N_c} & -{N_F \over 2}\\     
                                                N_c & 2N_c \\ \end{array} \right ).     
\label{m8g0}     
\eqnx     

\section{Unified treatment of the LO Altarelli Parisi evolution and of the       
double $ln^2(1/x)$ resummation.}       
     
In the region of large values of $x$ the integral equations        
(\ref{dlxns}), (\ref{dlxsg}) describing pure double logarithmic resummation      
$\ln^2(1/x)$, even completed by including non-ladder contributions,       
are inaccurate.      
In this region one should use the conventional Altarelli - Parisi  equations       
\cite{STRATMANN1,ALTAR,AP}      
with complete splitting functions $\Delta P_{ij}(z)$ and not restrict oneself      
to the effect generated only by their $z\rightarrow 0$ part.      
Following refs.\ \cite{BBJK,BZIAJA}, we do therefore extend       
equations (\ref{dlxns},\ref{dlxsg}) and add to their right hand side(s) the       
contributions coming from the      
remaining parts of splitting functions $\Delta P_{ij}(z)$.        
We also allow coupling $\alpha_S$ to run, setting $k^2$ as the relevant      
scale. In this way we obtain unified system of equations which contain      
both the complete leading order Altarelli - Parisi  evolution and the double       
logarithmic      
$ln^2(1/x)$ effects at low $x$.      
The corresponding system of equations reads~:      
{\small      
\eqn      
f_{NS}(x^{\prime},k^2)=f^{(0)}_{k}(x^{\prime},k^2)      
&+&      
{\alpha_S(k^2)\over 2 \pi}{4\over 3}      
\int_{x^{\prime}}^1 {dz\over z}      
\int_{k_0^2}^{k^2/z}      
{dk^{\prime 2}\over k^{\prime 2}}      
f_{NS}\left({x^{\prime}\over z},k^{\prime 2}\right)\label{unifns}\\      
&&\hspace*{10ex}{\bf (\hspace*{3ex}Ladder\hspace*{3ex})}\nonumber\\      
&+&{\alpha_S(k^2)\over 2\pi}\int_{k_0^2}^{k^2}{dk^{\prime 2}\over      
k^{\prime 2}}{4\over 3}\int _{x^{\prime}}^1      
{dz\over z} {(z+z^2)f_{NS}({x^{\prime}\over z},k^{\prime 2})-      
2zf_{NS}(x^{\prime},k^{\prime 2})\over 1-z}\nonumber\\      
&+&{\alpha_S(k^2)\over 2\pi}\int_{k_0^2}^{k^2}{dk^{\prime 2}\over      
k^{\prime 2}}\left[2 +      
{8\over 3} ln(1-x^{\prime})\right]f_{NS}(x^{\prime},k^{\prime 2})\nonumber\\      
&&\hspace*{10ex}{\bf(\hspace*{3ex}Altarelli - Parisi \hspace*{4ex})}\nonumber\\      
&-&{\alpha_S(k^2)\over 2\pi}      
\int_{x^{\prime}}^1 {dz\over z}      
\Biggl(      
\Biggl[ \frac{\tilde  {\bf F}_8 }{\omega^2} \Biggr](z)      
\frac{ {\bf G}_0 }{2\pi^2}      
\Biggr)_{qq}      
\int_{k_0^2}^{k^2}      
{dk^{\prime 2}\over k^{\prime 2}}      
f_{NS}\left({x^{\prime}\over z},k^{\prime 2}\right)\nonumber\\      
&-&{\alpha_S(k^2)\over 2\pi}      
\int_{x^{\prime}}^1 {dz\over z}      
\int_{k^2}^{k^2/z}      
{dk^{\prime 2}\over k^{\prime 2}}      
\Biggl(      
\Biggl[\frac{\tilde   {\bf F}_8 }{\omega^2} \Biggr]      
\Biggl(\frac{k^{\prime 2}}{k^2}z \Biggr)\frac{ {\bf G}_0 }{2\pi^2}      
\Biggr)_{qq}      
f_{NS}\left({x^{\prime}\over z},k^{\prime 2}\right),\nonumber\\      
&&\hspace*{10ex}{\bf(Non-ladder)}\nonumber      
\eqnx      
}      
{\small      
\eqn      
f_{S}(x^{\prime},k^2)=f^{(0)}_{S}(x^{\prime},k^2)      
&+&      
{\alpha_S(k^2)\over 2 \pi}      
\int_{x^{\prime}}^1 {dz\over z}      
\int_{k_0^2}^{k^2/z}      
{dk^{\prime 2}\over k^{\prime 2}}      
{4\over 3}      
f_{S}\left({x^{\prime}\over z},k^{\prime 2}\right)\nonumber\\      
&-&{\alpha_S(k^2)\over 2 \pi}\int_{x^{\prime}}^1 {dz\over z}      
\int_{k_0^2}^{k^2/z}      
N_F{dk^{\prime 2}\over k^{\prime 2}}f_{g}      
\left({x^{\prime}\over z},k^{\prime 2}\right)\nonumber\\      
&&\hspace*{10ex}{\bf (\hspace*{3ex}Ladder\hspace*{3ex})}\nonumber\\      
&+&{\alpha_S(k^2)\over 2 \pi}\int_{k_0^2}^{k^2}{dk^{\prime 2}\over      
k^{\prime 2}}{4\over 3}\int _{x^{\prime}}^1      
{dz\over z} {(z+z^2)f_{S}({x^{\prime}\over z},k^{\prime 2})-      
2zf_{S}(x^{\prime},k^{\prime 2})\over 1-z}\nonumber\\      
&+&{\alpha_S(k^2)\over 2 \pi}\int_{k_0^2}^{k^2}{dk^{\prime 2}\over      
k^{\prime 2}}\left[ 2 +      
{8\over 3} ln(1-x^{\prime})\right]      
f_{S}(x^{\prime},k^{\prime 2}) \nonumber\\      
&+&{\alpha_S(k^2)\over 2 \pi} \int_{k_0^2}^{k^2}{dk^{\prime 2}\over      
k^{\prime 2}}\int _{x^{\prime}}^1      
{dz\over z} 2z N_F f_g({x^{\prime}\over z},k^{\prime 2})\nonumber\\      
&&\hspace*{10ex}{\bf(\hspace*{3ex}Altarelli - Parisi \hspace*{4ex})}\nonumber\\      
&-&{\alpha_S(k^2)\over 2\pi}      
\int_{x^{\prime}}^1 {dz\over z}      
\Biggl(      
\Biggl[ \frac{\tilde  {\bf F}_8 }{\omega^2} \Biggr](z)      
\frac{ {\bf G}_0 }{2\pi^2}      
\Biggr)_{qq}      
\int_{k_0^2}^{k^2}      
{dk^{\prime 2}\over k^{\prime 2}}      
f_{S}\left({x^{\prime}\over z},k^{\prime 2}\right)\label{unifsea}\\      
&-&{\alpha_S(k^2)\over 2\pi}      
\int_{x^{\prime}}^1 {dz\over z}      
\int_{k^2}^{k^2/z}      
{dk^{\prime 2}\over k^{\prime 2}}      
\Biggl(      
\Biggl[\frac{\tilde   {\bf F}_8 }{\omega^2} \Biggr]      
\Biggl(\frac{k^{\prime 2}}{k^2}z \Biggr)\frac{ {\bf G}_0 }{2\pi^2}      
\Biggr)_{qg}      
f_{g}\left({x^{\prime}\over z},k^{\prime 2}\right),\nonumber\\      
&&\hspace*{10ex}{\bf(Non-ladder)}      
\nonumber      
\eqnx      
}      
{\small      
\eqn      
f_{g}(x^{\prime},k^2)=f^{(0)}_{g}(x^{\prime},k^2) &+&      
 {\alpha_S(k^2)\over 2 \pi}      
\int_{x^{\prime}}^1 {dz\over z}      
\int_{k_0^2}^{k^2/z}      
{dk^{\prime 2}\over k^{\prime 2}}      
{8\over 3}      
f_{S}\left({x^{\prime}\over z},k^{\prime 2}\right)\nonumber\\      
&+&{\alpha_S(k^2)\over 2 \pi}      
\int_{x^{\prime}}^1 {dz\over z}      
\int_{k_0^2}^{k^2/z}      
{dk^{\prime 2}\over k^{\prime 2}}      
12 f_{g}\left({x^{\prime}\over z},k^{\prime 2}\right)\nonumber\\      
&&\hspace*{10ex}{\bf (\hspace*{3ex}Ladder\hspace*{3ex})}\nonumber\\      
&+&{\alpha_S(k^2)\over 2 \pi}      
\int_{k_0^2}^{k^2}      
{dk^{\prime 2}\over k^{\prime 2}}\int_{x^{\prime}}^1 {dz\over z}      
(-{4\over 3})zf_{S}      
\left({x^{\prime}\over z},k^{\prime 2}\right)\nonumber\\      
&+&{\alpha_S(k^2)\over 2 \pi}      
\int_{k_0^2}^{k^2}      
{dk^{\prime 2}\over k^{\prime 2}} \int_{x^{\prime}}^1 {dz\over z}      
6z\left[{f_{g}      
\left({x^{\prime}\over z},k^{\prime 2}\right)- f_{g}      
(x^{\prime},k^{\prime 2})\over 1-z} -2f_{g}      
\left({x^{\prime}\over z},k^{\prime 2}\right)\right]\nonumber\\      
&+&{\alpha_S(k^2)\over 2 \pi}      
\int_{k_0^2}^{k^2}      
{dk^{\prime 2}\over k^{\prime 2}}\left[ {11\over 2} -{N_F\over 3}      
 + 6 ln(1-x^{\prime})\right]f_{g}      
(x^{\prime},k^{\prime 2})\nonumber\\      
&&\hspace*{10ex}{\bf(\hspace*{3ex}Altarelli - Parisi \hspace*{4ex})}\nonumber\\      
&-&{\alpha_S(k^2)\over 2\pi}      
\int_{x^{\prime}}^1 {dz\over z}      
\Biggl(      
\Biggl[ \frac{\tilde  {\bf F}_8 }{\omega^2} \Biggr](z)      
\frac{ {\bf G}_0 }{2\pi^2}      
\Biggr)_{gq}      
\int_{k_0^2}^{k^2}      
{dk^{\prime 2}\over k^{\prime 2}}      
f_{S}\left({x^{\prime}\over z},k^{\prime 2}\right)\label{unifglue}\\      
&-&{\alpha_S(k^2)\over 2\pi}      
\int_{x^{\prime}}^1 {dz\over z}      
\int_{k^2}^{k^2/z}      
{dk^{\prime 2}\over k^{\prime 2}}      
\Biggl(      
\Biggl[\frac{\tilde   {\bf F}_8 }{\omega^2} \Biggr]      
\Biggl(\frac{k^{\prime 2}}{k^2}z \Biggr)\frac{ {\bf G}_0 }{2\pi^2}      
\Biggr)_{gg}      
f_{g}\left({x^{\prime}\over z},k^{\prime 2}\right).\nonumber\\      
&&\hspace*{10ex}{\bf(Non-ladder)}\nonumber      
\eqnx      
}      
In equations (\ref{unifns}), (\ref{unifsea}), (\ref{unifglue})      
we group separately terms corresponding to ladder      
diagram contributions to the double $ln^2(1/x)$ resummation,      
contributions from the non-singular parts of the Altarelli - Parisi       
splitting functions and finally contributions from the non-ladder      
bremsstrahlung diagrams.  We label those three contributions as      
"ladder", "Altarelli - Parisi " and "non-ladder" respectively.      
      
Inhomogeneous terms $f_i^{(0)}(x^{\prime},k^2)$ ($i=NS,S, g$),      
as stated above, may be expressed as~:      
\eqn      
f_{NS}^{(0)}(x^{\prime},k^2)&=&k^{\gamma}_{NS}(x)+      
{\alpha_S(k^2)\over 2 \pi}{4\over 3}      
\int _{x^{\prime}}^1      
{dz\over z} {(1+z^2)\Delta q_{NS}^{(0)}({x^{\prime}\over z})-      
2z\Delta q_{NS}^{(0)}(x^{\prime})\over 1-z}\nonumber\\      
&&+{\alpha_S(k^2)\over 2 \pi}\left[2 +{8\over 3}      
ln(1-x^{\prime})\right]\Delta q_{NS}^{(0)}(x^{\prime})      
\label{fns0},      
\eqnx      
\eqn      
f_{S}^{(0)}(x^{\prime},k^2) &=&k^{\gamma}_{S}(x)      
+{\alpha_S(k^2)\over 2 \pi}{4\over 3}      
\int _{x^{\prime}}^1      
{dz\over z} {(1+z^2)\Delta q_S^{(0)}({x^{\prime}\over z})-      
2z\Delta q_S^{(0)}(x^{\prime})\over 1-z}\nonumber\\      
&&+{\alpha_S(k^2)\over 2 \pi}(2 +{8\over 3} ln(1-x^{\prime}))      
\Delta q_S^{(0)}(x^{\prime})\nonumber\\      
&&+{\alpha_S(k^2)\over 2 \pi}N_F\int _{x^{\prime}}^1{dz\over z}      
(1-2z)\Delta g^{(0)}({x^{\prime}\over z}),\nonumber\\      
f_g^{(0)}(x^{\prime},k^2) &=&k^{\gamma}_{g}(x)      
+{\alpha_S(k^2)\over 2 \pi}{4\over 3}      
\int _{x^{\prime}}^1{dz\over z}(2-z)\Delta q_S^{(0)}({x^{\prime}\over      
z})\nonumber\\      
&&+{\alpha_S(k^2)\over 2 \pi}({11\over 2} -{N_F\over 3}      
 + 6 ln(1-x^{\prime}))\Delta g^{(0)}(x^{\prime})\label{fsg0}\\      
&&+{\alpha_S(k^2)\over 2 \pi}6      
\int_{x^{\prime}}^1 {dz\over z}\left[      
{\Delta g^{(0)}      
({x^{\prime}\over z})- z\Delta g^{(0)}      
(x^{\prime})\over 1-z} +(1-2z)\Delta g^{(0)}      
({x^{\prime}\over z})\right].\nonumber      
\eqnx      
Equations (\ref{unifns}), (\ref{unifsea}), (\ref{unifglue}) together with      
(\ref{fns0}), (\ref{fsg0}) and      
(\ref{dpi}) reduce to the leading order Altarelli - Parisi  evolution equations for      
photonic structure function with starting (integrated) distributions      
$\Delta q_i^{(0)}(x)$ ($i=NS,S$) and $\Delta g^{(0)}(x)$ after we set the upper      
integration limit over $dk^{\prime 2}$ equal to $k^2$ in all terms in      
equations (\ref{unifns}), (\ref{unifsea}), (\ref{unifglue}),      
neglect the higher order terms in the kernels,      
and set $Q^2$ in place of $W^2$ as the upper integration limit of the integral      
in eq.\ (\ref{dpi}).   In this approximation the cut-off parameter $k_0^2$    
is  equal to the magnitude of the scale $Q^2$ for which    
the parton distributions in the photon are entirely specified by the hadronic component    
of the photon and are equal to    
$\Delta q_i^{(0)}(x)$ ($i=NS,S$) and $\Delta g^{(0)}(x)$.

\section{Numerical results.}      
\begin{figure}[t]      
\centerline{\epsfig{figure=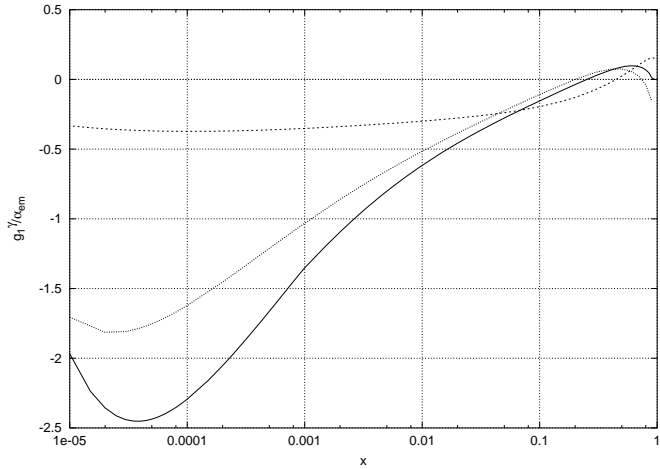,width=9cm}}      
\caption{{\footnotesize Structure function $\gp(x,Q^2)/\alpha_{em}$     
for $Q^2=10 GeV^2$ derived      
after solving eqs.\ (\ref{unifns}), (\ref{unifsea}), (\ref{unifglue})        
with input parametrization (\ref{par1}) plotted as function of $x$.      
Solid line corresponds to the  calculations which contain full      
$ln^2(1/x)$ resummation with both bremsstrahlung corrections and LO Altarelli - Parisi       
kernel included, dashed line shows pure LO Altarelli - Parisi  evolution,  
dotted line represents NLO Altarelli - Parisi  evolution.}}      
\label{fig.1}      
\end{figure}      
%
\begin{figure}[t]      
\centerline{\epsfig{figure=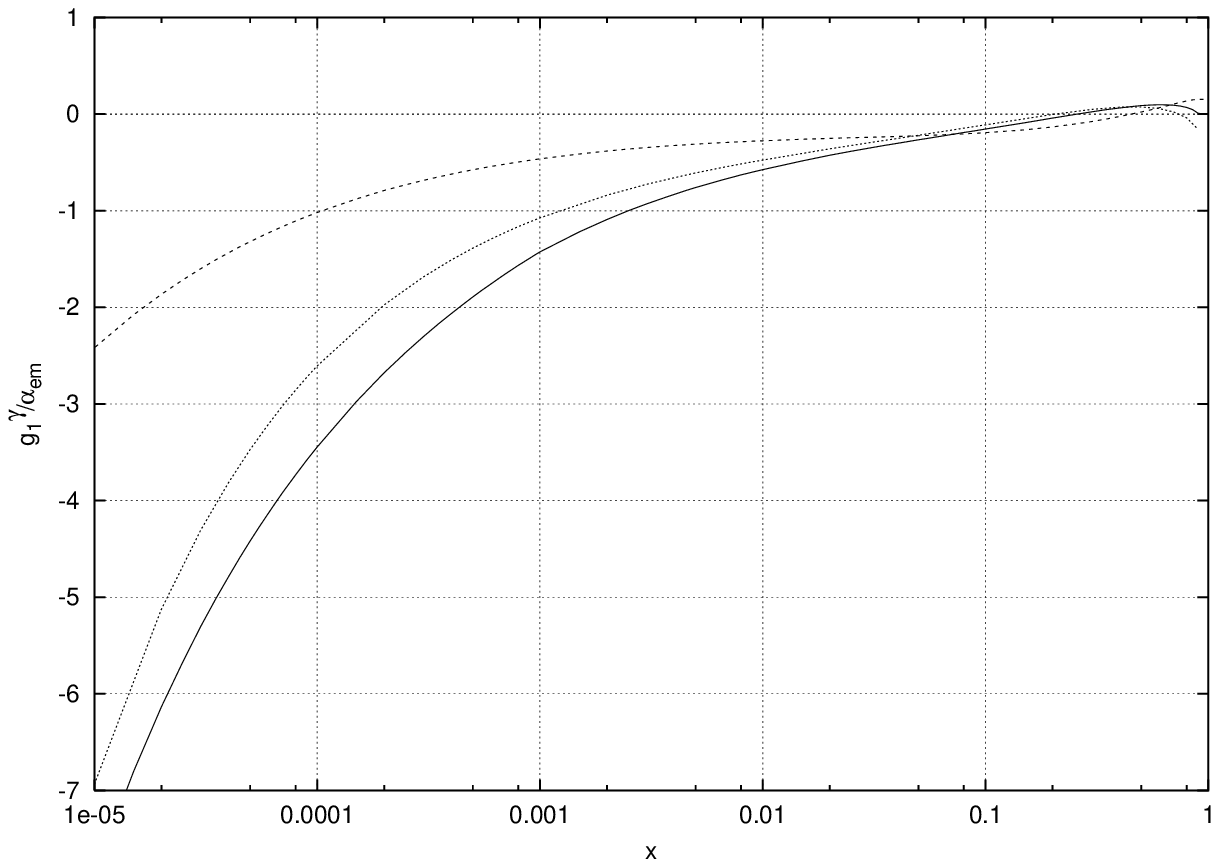,width=9cm}}      
\caption{{\footnotesize Structure function $\gp(x,Q^2)/\alpha_{em}$ for $Q^2=10 GeV^2$ derived      
after solving eqs.\ (\ref{unifns}), (\ref{unifsea}), (\ref{unifglue})        
with input parametrization (\ref{par20}) plotted as function of $x$.      
Solid line corresponds to the  calculations which contain full      
$ln^2(1/x)$ resummation with both bremsstrahlung corrections and LO Altarelli - Parisi       
kernel included, dashed line shows  pure LO Altarelli - Parisi       
evolution, dotted line represents NLO Altarelli - Parisi  evolution.}}      
\label{fig.1}      
\end{figure}      
%
\begin{figure}[t]      
\centerline{\epsfig{figure=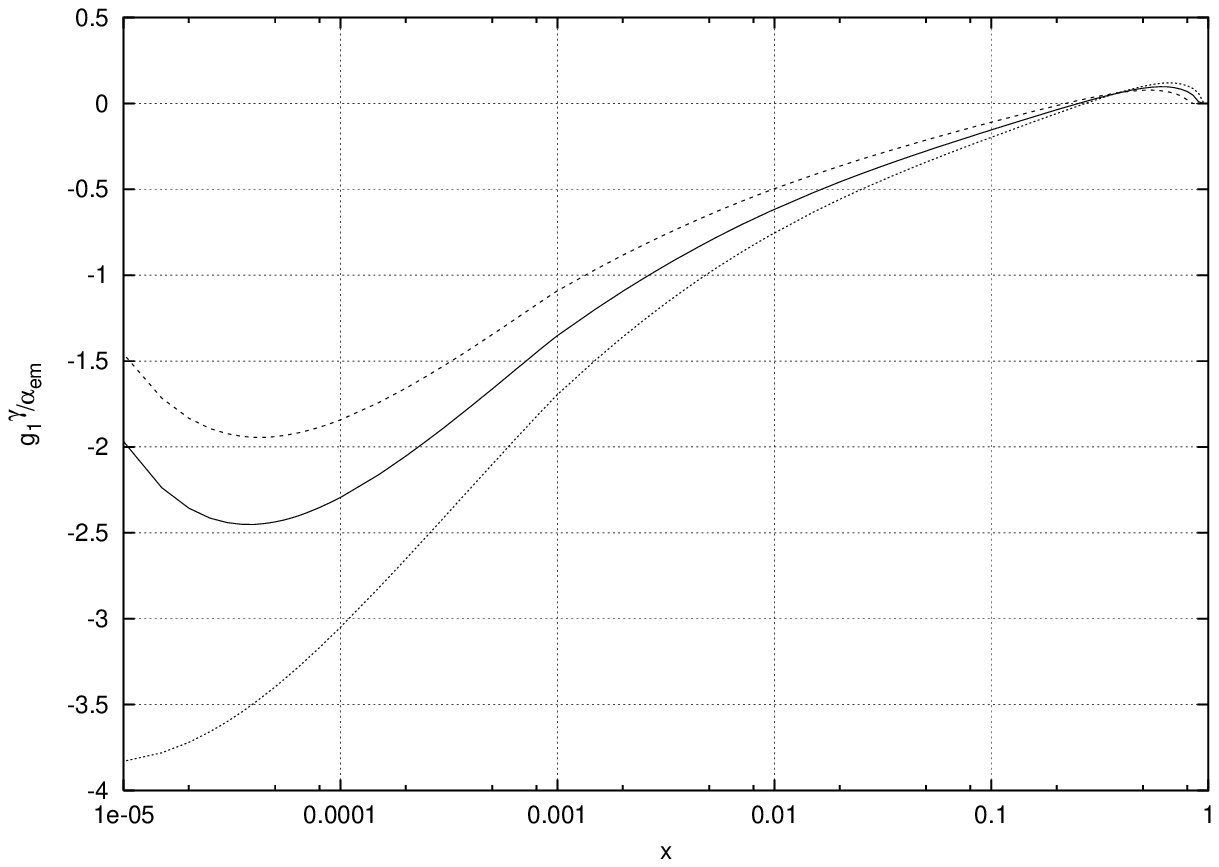,width=9cm}}      
\caption{{\footnotesize Sensitivity of predictions for $g_1^{\gamma}(x,Q^2)$   
($Q^2=10 GeV^2$) with input parametrization (\ref{par1}) to the magnitude of   
the infrared cut-off  parameter $k_0^2$. Solid line corresponds to the   
calculations with $k_0^2=1\,GeV^2$,   
dashed line shows $g_1^{\gamma}(x,Q^2)$ for $k_0^2=2.0\,GeV^2$ and  
dotted line represents $g_1^{\gamma}(x,Q^2)$ calculations   
for $k_0^2=0.5\,GeV^2$.}}      
\label{fig.1}      
\end{figure}      
%
\begin{figure}[t]      
\centerline{\epsfig{figure=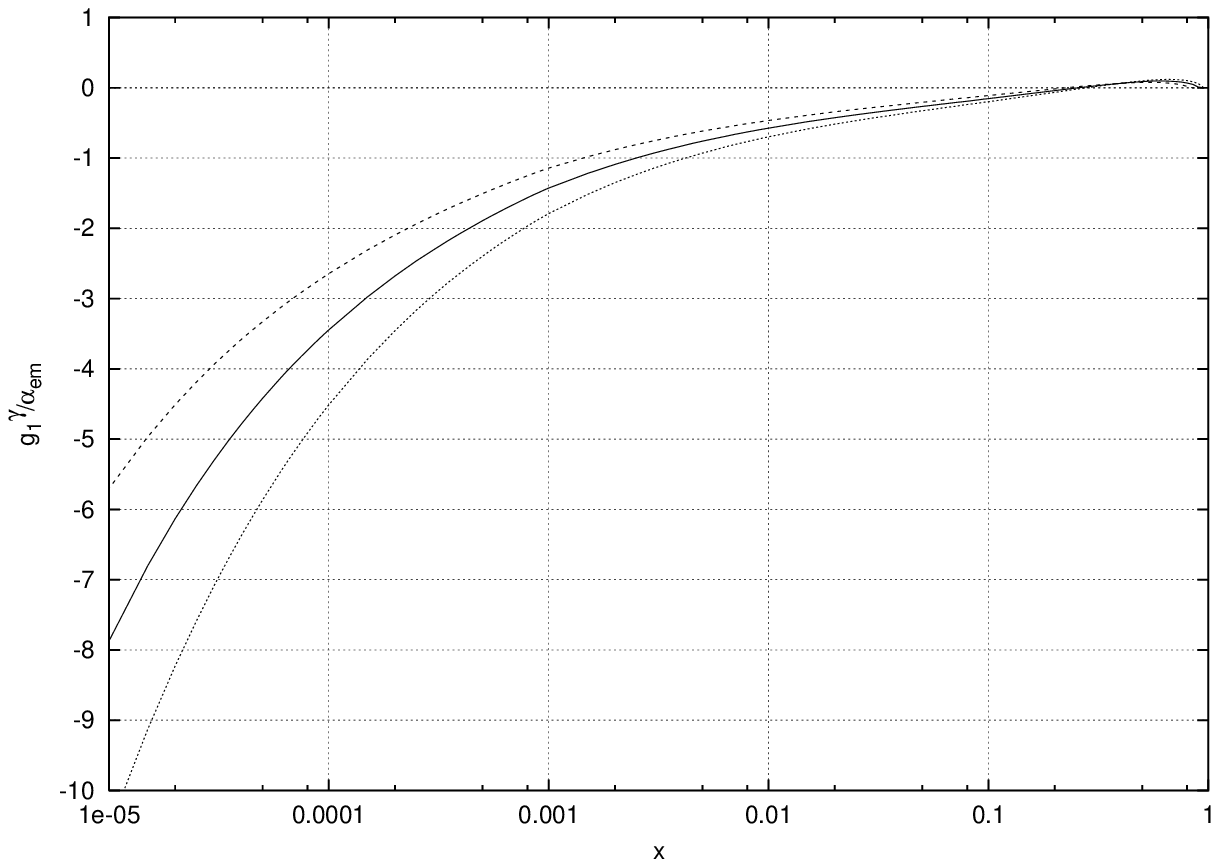,width=9cm}}      
\caption{{\footnotesize Sensitivity of predictions for $g_1^{\gamma}(x,Q^2)$   
($Q^2=10 GeV^2$) with input parametrization (\ref{par20}) to the magnitude of the infrared   
cut-off  parameter $k_0^2$. Solid line corresponds to the calculations with   
$k_0^2=1\,GeV^2$,   
dashed line shows $g_1^{\gamma}(x,Q^2)$ for $k_0^2=2.0\,GeV^2$ and  
dotted line represents $g_1^{\gamma}(x,Q^2)$ calculations   
for $k_0^2=0.5\,GeV^2$.}}      
\label{fig.1}      
\end{figure}      
%
\begin{figure}[t]      
\centerline{\epsfig{figure=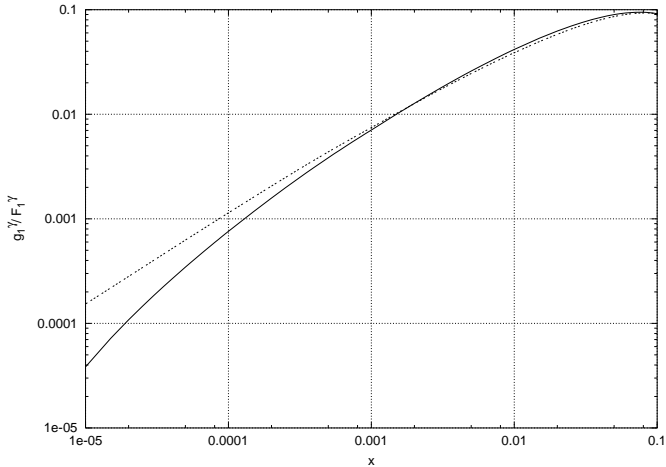,width=9cm}}      
\caption{{\footnotesize Absolute value of asymmetry 
$g_1^{\gamma}(x,Q^2)/F_1^{\gamma}(x,Q^2)$ at small $x$  
($Q^2=10 GeV^2$) plotted as function of $x$. Solid line corresponds to the    
calculations with input parametrization (\ref{par1}), dashed line represents  
asymmetry for input parametrization (\ref{par20}).}}      
\label{fig.1}      
\end{figure}      
%
%
\begin{figure}[t]      
\centerline{\epsfig{figure=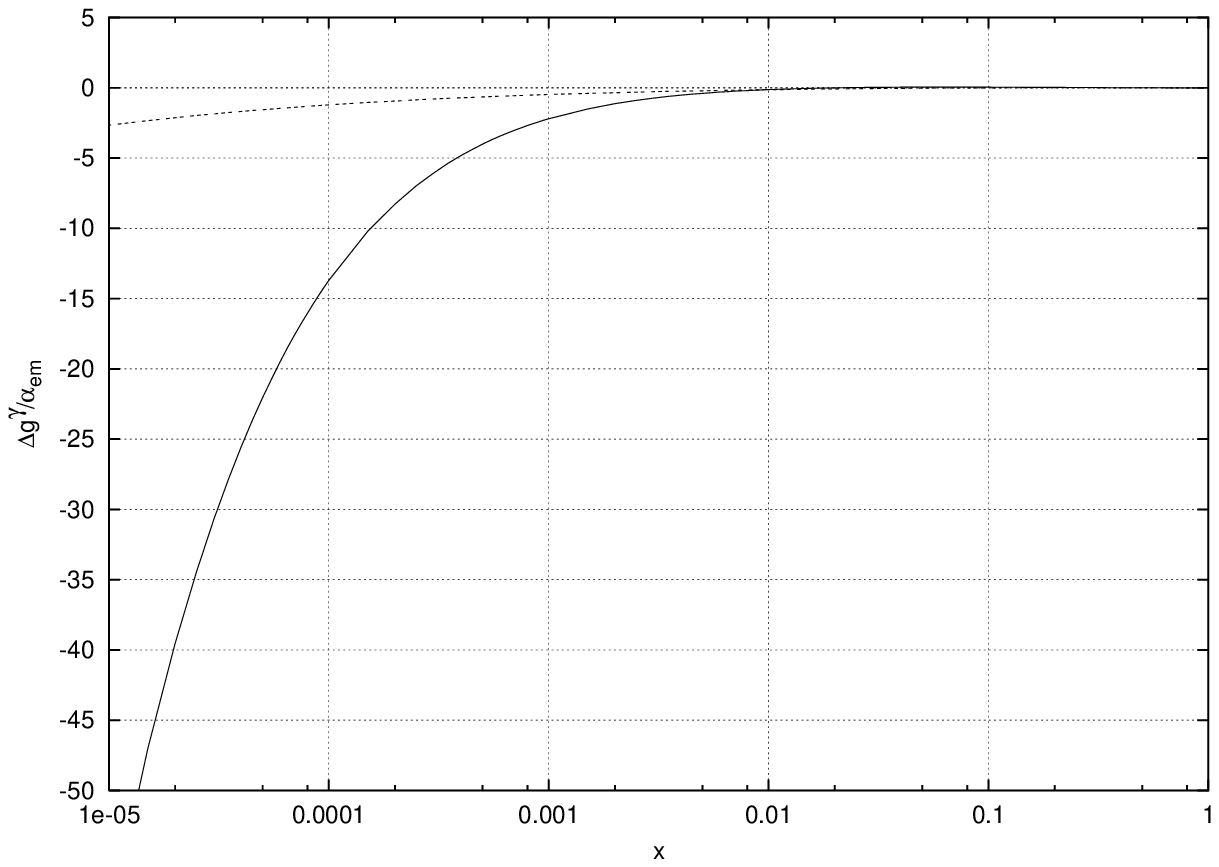,width=9cm}}      
\caption{{\footnotesize Spin dependent gluon distribution in the photon    
$\Delta g^{\gamma}(x,Q^2)/\alpha_{em}$       
for $Q^2=10 GeV^2$ derived after solving eqs.\ (\ref{unifns}), (\ref{unifsea}), (\ref{unifglue})        
with input parametrization (\ref{par1}) plotted as function of $x$.      
Solid line corresponds to the  calculations which contain full      
$ln^2(1/x)$ resummation with both bremsstrahlung corrections and LO Altarelli - Parisi       
kernel included, dashed line shows  pure LO Altarelli - Parisi       
evolution.}}      
\label{fig.1}      
\end{figure}      
%
\begin{figure}[t]      
\centerline{\epsfig{figure=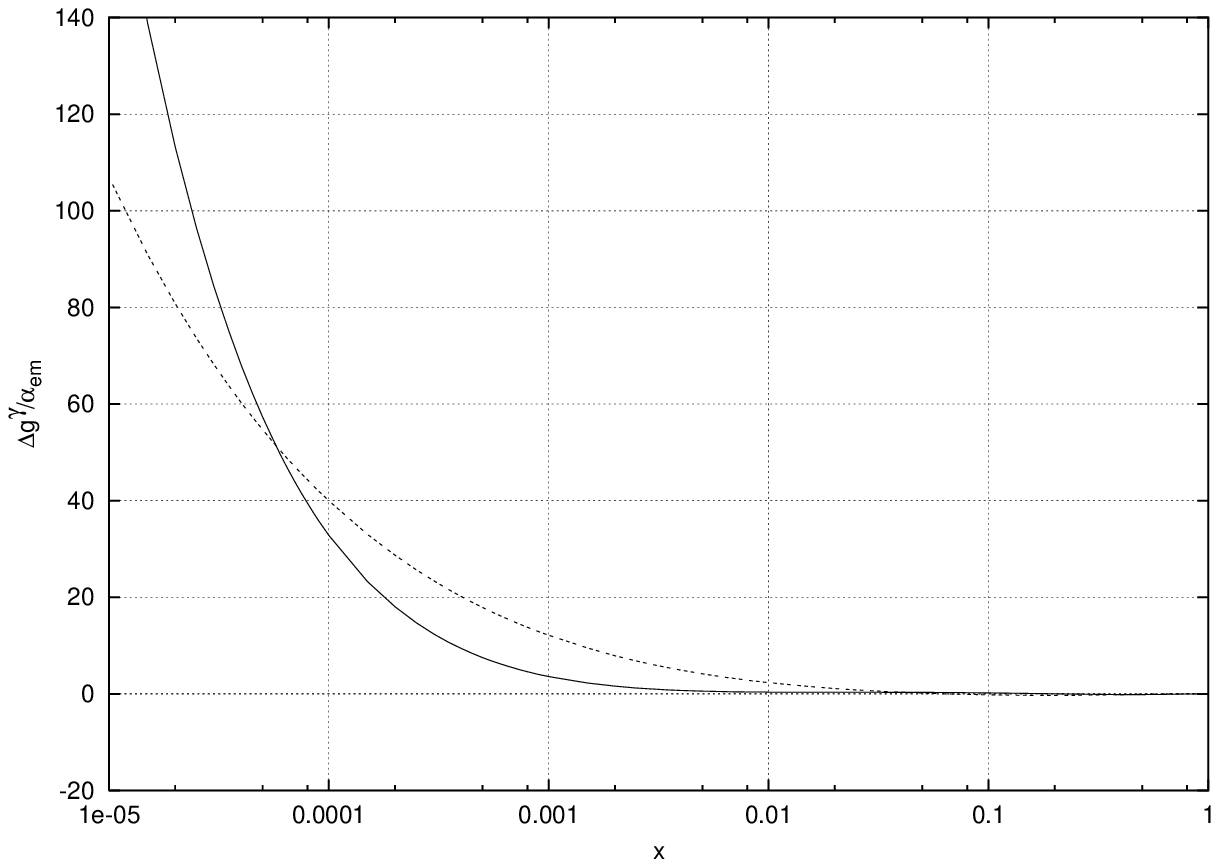,width=9cm}}      
\caption{{\footnotesize Spin dependent gluon distribution    
$\Delta g(x,Q^2)/\alpha_{em}$       
for $Q^2=10 GeV^2$ derived after solving eqs.\ (\ref{unifns}), (\ref{unifsea}), (\ref{unifglue})        
with input parametrization (\ref{par20}) plotted as function of $x$.      
Solid line corresponds to the  calculations which contain full      
$ln^2(1/x)$ resummation with both bremsstrahlung corrections and LO Altarelli - Parisi       
kernel included, dashed line shows  pure LO Altarelli - Parisi       
evolution.}}      
\label{fig.1}      
\end{figure}      

We have solved equations (\ref{unifns}), (\ref{unifsea}), (\ref{unifglue}),      
assuming the input~parametrizations which satisfy the constraint      
for the first moment of photonic structure function $\gp(x,Q^2)$ \cite{SUMR}:       
\eq      
\int_{0}^1 \gp(x,Q^2)=0      
\label{g1}      
\eqx      
which, in turn, implies the non-perturbative input to fulfill the      
sum rule ($l=NS,S$)~:      
\eqn      
\int_{0}^1 \Delta q_l^{(0)} (x)&=&0\nonumber,\\    
\int_{0}^1 \Delta g^{(0)} (x)&=&0.     
\label{dp1}      
\eqnx      
We have considered two input parametrizations and tested their influence on the      
behaviour of $\gp(x,Q^2)$. The first parametrization assumed no non-perturbative      
contribution at all and followed the lower limit parametrization by Stratmann      
et al.\ \cite{STRATMANN1,STRATMANNL}:       
\eqn      
\Delta q_l^{(0)} (x)&=&0\nonumber,\\    
\Delta g^{(0)} (x)&=&0.      
\label{par1}      
\eqnx      
In the  second parametrization we allowed for the non-zero input       
gluon distribution which was based  on the vector meson dominance (VMD)       
model. Assuming the dominance of the $\rho$ and $\omega$      
meson contribution, one obtains~:       
\eq      
\Delta g^{(0)}(x)=\alpha_{em}\left(g_{\rho}^2\,\Delta g^{\rho\,(0)}(x)      
+g_{\omega}^2\,\Delta g^{\omega\,(0)}\right),      
\label{par20}      
\eqx      
where $g_{\rho}$  and $g_{\omega}$  are  constants characterizing the coupling      
of the $\rho$ and $\omega$ mesons to the photon, and:     
\eq     
g_{\rho}^2+g_{\omega}^2\sim 0.5.     
\label{rhoom}     
\eqx     
Coefficient $\Delta g^{V\,(0)}(x)$ ($V=\rho,\omega$) denotes the nonperturbative part of the       
spin dependent gluon distribution in a photon which was parametrized in the       
following form:      
\eq      
\Delta g^{V\,(0)}(x)=C_g^{V}\,(1-x)^{3}\,(1+a_g\,x),      
\label{par21}      
\eqx      
so that we have $\Delta g^{V\,(0)}(x) \rightarrow C_g^{V}$ in the limit       
$x \rightarrow 0$.  Similar behaviour was assumed in ref. \cite{BZIAJA}       
for the nonperturbative part of the spin dependent gluon distribution in       
the nucleon, i.e. $\Delta g^{N\,(0)}(x) \rightarrow C_g^{N}$.     
Constant $C_g^{V}$ was related to constant $C_g^{N}$ as below       
\eq      
C_l^V=\frac{2}{3}\,C_l^N,     
\label{c}      
\eqx      
and it reads $C_g^{V}=4.5$. Finally, coefficient $a_l$ is determined by imposing       
condition (\ref{dp1}) on the distributions (\ref{par20}) so that      
$a_l=-5$.      
   
In Figs.\ 2, 3 we present results obtained for photonic structure function      
$\gp(x,Q^2)$ after numerical solving of eqs.\ (\ref{unifns}),      
(\ref{unifsea}), (\ref{unifglue}) with parametrizations (\ref{par1}),      
(\ref{par20}) respectively.      
      
In Fig.\ 2 predictions for $\gp(x,Q^2)$ based on eqs.\ (\ref{unifns}),      
(\ref{unifsea}), (\ref{unifglue}) with input parametrization (\ref{par1})      
are plotted and confronted with the results obtained from the solution      
of LO and NLO Altarelli - Parisi  evolution equations with the input    
distribution given by      
eq.\ (\ref{par1}). One may see that double logarithmic contributions      
manifest at low $x$, however, not as strong as for the nucleon structure      
function case (cf. \cite{BZIAJA}). There is a characteristic decrease of photonic      
structure function till $x\sim 10^{-4}$, then the function starts to      
increase rapidly,      
and it is expected to grow steeply in the ultra-asymptotic region $x<10^{-5}$.      
We note that in the small $x$ region which may be probed in linear $e \gamma$       
colliders (i.e. $x > 10^{-4}$) the leading asymptotic behaviour       
has not been established yet, and that it is delayed to the ultra-asymptotic       
region.  This is  manifestation of the effect  discussed in Sec. 2.        
      
In Fig.\ 3 we show results for $\gp(x,Q^2)$ which were obtained assuming       
the presence of the non-perturbative part of the spin dependent gluon distribution       
in the  photon (see eq.\ (\ref{par20})). One may see from this plot     
that the  absolute magnitude of the structure function is significantly enhanced     
in comparison to the previous       
case.  This is related to the enhancement of the leading asymptotic       
term by the non-perturbative gluon distribution.       
The structure function would stay negative in the limit $x \rightarrow 0$.\\   
      
In both cases we notice that    
the NLO approximation of the Altarelli-Parisi equation    
contains significant part of the double $ln^2(1/x)$ resummation in the structure    
function $g_1(x,Q^2)$ at low values of $x$.  It should be emphasised that this resummation    
embodies the leading singular part of the NLO  approximation which contains     
the lowest order term of the double $ln^2(1/x)$    
resummation. We also notice that the difference between the structure function    
$g_1(x,Q^2)$ calculated from equations (\ref{dpi}), (\ref{g11}), (\ref{unifns}),  
(\ref{unifsea}), (\ref{unifglue}) and that obtained within the LO  Altarelli-Parisi    
formalism is also present at large values of $x$.  This fact is caused by    
different upper  integration limit in eq.\ (\ref{dpi}) which    
for the LO approximation is equal to $Q^2$ instead of $W^2=Q^2(1/x-1)$.     
Although this difference is formally subleading it is logarithmically    
enhanced  at both small and large values of $x$.  At large values of $x$    
the unintegrated parton distributions $f_l(x,k^2)$ are dominated    
by the inhomogeneous parts $k_l^{\gamma}(x)$ (see eq.\ (\ref{ftok})), and    
we get $g_1^{\gamma}(x,Q^2)\sim \kappa(x) ln(W^2/k_0^2)$    
(modulo less singular terms)    
while in LO approximation  we have $g_1^{\gamma}(x,Q^2)\sim \kappa(x) ln(Q^2/k_0^2)$, where    
the function $\kappa(x)$ is defined by eq.\ (\ref{kappa}).  We do therefore notice    
that structure function $g_1^{\gamma}(x,Q^2)$ calculated within our scheme    
contains at large values of $x$ an additional term proportional to    
$\kappa(x)ln(1/x-1)$ in comparison    
to the structure function calculated within the LO approximation, since    
$ln(W^2/k_0^2) = ln(Q^2/k_0^2) + ln(1/x-1)$. This additional     
term is in    
fact equal to the singular part of the photonic coefficient    
$\Delta C_{\gamma}(x)$  in the NLO    
approximation in the $\overline{MS}$ scheme \cite{STRATMANN1}.     
These effects are of course absent in the hadronic case and so the    
spin dependent nucleon structure    
function calculated within the formalism generating the double $ln^2(1/x)$    
resummation approaches at large values of $x$ that calculated within the LO    
approximation \cite{BZIAJA}.\\

In Figs.\ 4, 5 we show sensitivity of our predictions for $g_1^{\gamma}(x,Q^2)$ at low    
values of $x$ to the    
magnitude of the infrared cut-off  parameter $k_0^2$.  We see that the    
magnitude of the structure function can change  by about a factor equal to 1.8 for    
$k_0^2$ varying between $0.5$ and $2$~$GeV^2$.  The cut-off parameter    
$k_0^2$ does also slightly affect the shape of $g_1^{\gamma}(x,Q^2)$ as the function    
of $x$.\\   
   
In Fig.\ 6 we plot the ratio $g_1^{\gamma}(x,Q^2)/F_1^{\gamma}(x,Q^2)$ 
which measures the asymmetry, using the structure function $F_1^{\gamma}(x,Q^2)$
obtained from the LO analysis  presented in ref. \cite{GRS}.    
We see that the magnitude of this ratio    
is very small for $x < 10^{-3}$ (i.e. smaller than $10^{-2}$), and it may    
be very difficult to measure such a low value of the asymmetry.\\

We have also plotted spin dependent photonic gluon      
distribution $\Delta g^{\gamma}$ derived for both parametrizations (\ref{par1}),      
(\ref{par20}) (see Figs.\ 7, 8 ). We observe that the gluon distributions       
are very different.   We note in particular that $\Delta g^{\gamma}(x,Q^2)$       
is negative at small $x$ for the first case, i.e. when this distribution       
is generated purely radiatively.  It is possible to obtain a positive       
gluon  distribution assuming the (positive) non-perturbative part of this       
distribution (see Fig. 8). These two possible scenarios for the       
gluon distributions could be discriminated by the measurements which       
would access more directly  the spin dependent gluon distributions       
in the photon like, for instance, the measurement of the dijet production       
in (polarised) $\gamma \gamma$ scattering etc.         
     
\section{Summary and conclusions}     
      
In this paper we have studied the possible effects of the double       
$ln^2(1/x)$ resummation upon behaviour of spin dependent       
structure function $g_1^{\gamma}(x,Q^2)$ of the photon.  This quantity       
could in principle be measured in the future linear colliders, in particular       
in the $e \gamma$ mode. We have extended the formalism developed by us for the       
case of the spin dependent structure function of the nucleon to the       
case of the photon structure functions by including the suitable inhomogeneus       
terms in the corresponding integral equations.  These inhomogeneous terms       
describe the point-like coupling of the (real) photons to quarks       
and antiquarks.  We have studied sensitivity  of our predictions upon the       
possible hadronic component of the spin dependent distributions, and found that       
the presence of the (input) gluon distribution in the hadronic component       
can significantly enhance the absolute magnitude of the structure function      
in the low $x$ region.         
      
\section*{Acknowledgments}      
      
We thank Gunnar Ingelman for reading the manuscript and useful comments.      
This research has been supported in part by the Polish Committee for Scientific      
Research grant 2 P03B 04718 and European Community grant      
'Training and Mobility of Researchers', Network 'Quantum     
Chromodynamics and the Deep Structure of Elementary Particles'     
FMRX-CT98-0194. B.\ Z.\ is a fellow of Stefan Batory Foundation.     


\begin{thebibliography}{9999}      
%
\bibitem{BARTNS}J. Bartels, B.I. Ermolaev and M.G. Ryskin,      
Z. Phys. {\bf C70} (1996) 273.      
%
\bibitem{BARTS}J. Bartels, B.I. Ermolaev and M.G. Ryskin, Z. Phys.      
{\bf C72} (1996) 627.      
%
\bibitem{RG} J. Bl\"umlein and A. Vogt, Acta Phys. Polon. {\bf B27} (1996) 1309;      
 Phys.Lett. {\bf B386} (1996) 350; J. Bl\"umlein, S. Riemersma and A. Vogt      
 Nucl. Phys. B (Proc. Suppl)  {\bf 51C} (1996) 30.      
%
\bibitem{JAP} Y.\ Kiyo, J.\ Kodaira, H.\ Tochimura, Z.\ Phys.\ {\bf C74}      
(1997) 631-639.      
%
\bibitem{BBJK}B. Bade\l{}ek, J. Kwieci\'nski,      
 Phys. Lett. {\bf B418} (1998) 229.      
%
\bibitem{BZIAJA}      
J. Kwieci\'nski, B. Ziaja, Phys. Rev. {\bf D60} (1999)  054004.       
%
\bibitem{ALBERT}      
A. De Roeck, Proceedings of the Cracow Epiphany Conference on Spin Effects in Particle      
Physics, Krak\'ow, Poland, January 9-11, 1998, Acta Phys. Polon. {\bf B29} (1998)      
1343.      
%
\bibitem{DESYW} Proceedings of the workshop 'Physics with Polarized Protons at HERA',       
DESY, 1997, eds. A. De Roeck and T. Gehrmann, DESY-PROCEEDINGS-1998-01.       
%
\bibitem{STRATMANN1} M. Stratmann, W. Vogelsang, Phys. Lett. {\bf B386}     
(1996) 370.       
%
\bibitem{STRATMANNL}M. Stratmann, Proceedings of the conference 'Lund 1998,     
Photon interactions and the photon structure' 135.      
%
\bibitem{STRATMANN2} M. Stratmann, Nucl. Phys. Proc. Suppl. {\bf 82} (2000)     
400.     
%
\bibitem{STRATMANN3}M. Stratmann and W. Vogelsang, Proceedings of the conference     
' Hamburg 1999. Polarized protons at high energies - Accelerator challenges      
and physics opportunities' 324.     
%
\bibitem{AVOGT} A. Vogt, Nucl. Phys. Proc. Suppl. {\bf 82} (2000)     
394.     
%
\bibitem{GAMMAC}I.F. Ginzburg, Nucl. Inst. Meth. {\bf 205} (1983) 47;     
{\it ibid}     
{\bf 219} (1984) 5; V.I. Telnov, Nucl. Inst. Meth. {\bf A294} 72;       
D.I. Borden, D.A. Bauer, and D.O. Caldwell, Phys. Rev. {\bf D48} (1993) 4018.      
\bibitem{MAN}  S.I. Manayenkov, M.G. Ryskin,  Proceedings of the Workshop "      
Physics with Polarized Protons at HERA", DESY March-Septmber 1997, edited      
by~: A. De Roeck and T. Gehrmann, DESY PROCEEDINGS-1998-01.      
%
\bibitem{QCD} R. Kirschner and L.N. Lipatov, Nucl. Phys. {\bf      
B213} (1983) 122; R. Kirschner, Z. Phys. {\bf C67} (1995) 459.      
\bibitem{QED} V.G. Gorshkov {\it et al.}, Yad. Fiz. {\bf 6} (1967)      
129 (Sov. J. Nucl. Phys. {\bf 6} (1968) 95); L.N. Lipatov, Zh.      
Eksp. Teor. Fiz. {\bf 54} (1968) 1520 (Sov. Phys. JETP {\bf 27}      
(1968) 814.      
%
%
\bibitem{ALTAR}      
G. Altarelli {\it et al.}, Proceedings of the Cracow Epiphany Conference on Spin Effects in Particle      
Physics, Krak\'ow, Poland, January 9-11, 1998, Acta Phys. Polon. {\bf B29} (1998)      
1145.      
\bibitem{AP}M. Ahmed and G. Ross, Phys. Lett. {\bf B56} (1976)      
385; Nucl. Phys. {\bf 111} (1976) 298; G. Altarelli and G. Parisi, Nucl. Phys. {\bf B126}      
(1977) 298; M. Stratmann, A. Weber and W. Vogelsang,      
Phys. Rev.{\bf D53} (1996) 138; M. Gl\"uck, {\it et al.}       
Phys. Rev. {\bf D53} (1996) 4775;  M. Gl\"uck, E. Reya and W. Vogelsang,      
Phys. Lett. {\bf B359} (1995) 201; T. Gehrmann and W. J. Stirling,      
Phys. Rev. {\bf D53} (1996) 6100; Phys. Lett. {\bf B365} (1996) 347;      
C. Bourrely and J. Soffer, Nucl. Phys. {\bf B445} (1995) 341;      
Phys. Rev. {\bf D53} (1996) 4067; R. D. Ball, S. Forte and G. Ridolfi,      
Nucl. Phys. {\bf B444} (1995) 287;      
J. Bartelski and S. Tatur, Acta Phys. Polon. {\bf B27}      
(1996) 911; Z. Phys. {\bf C71} (1996) 595;  E. Leader, A. V. Sidorov,     
D. B. Stamenov,     
Int.  J. Mod. Phys.  {\bf A13} (1998) 5573;  Phys. Lett. {\bf B445} (1998) 232.      
%
\bibitem{SUMR} S.D. Bass, Int. J. Mod. Phys. {\bf A7} (1992) 6039;       
S. Narison, G.M. Shore, and G. Veneziano, Nucl. Phys.       
{\bf B391} (1993) 69;       
S.D. Bass, S.J. Brodsky, and I. Schmidt, Phys. Lett. {\bf B437}       
 (1998) 417.      
\bibitem{GRS} M. Gl\"uck, E. Reya, I. Schienbein,  Phys. Rev. D60 (1999) 054019;    
 Erratum-ibid. D62 (2000)   
     019902.    
\end{thebibliography}
\end{document}